\documentclass[natbib]{elsarticle}

\usepackage{graphicx}
\usepackage{dcolumn}
\usepackage{booktabs}
\usepackage{bm}
\usepackage{graphicx}
\usepackage{breakurl}
\usepackage{graphicx}
\usepackage{amsmath, amssymb}
\usepackage{hyperref}

\newcommand{\msun}{M_{\odot}}

\begin{document}

\title{Constraining Mirror Dark Matter Inside Neutron Stars}

\author{Raul Ciancarella and Francesco Pannarale}
\ead{raul.cianca@gmail.com}
\address{Dipartimento di Fisica, Universit\`a di Roma ``Sapienza'', Piazzale A.~Moro 5, I-00185, Rome, Italy}
\address{INFN, Sezione di Roma, Piazzale A.~Moro 5, I-00185, Rome, Italy}

\author{Andrea Addazi}
\ead{addazi@scu.edu.cn}
\address{Center for Theoretical Physics, College of Physics Science and Technology, Sichuan University, 610065 Chengdu, China}
\address{INFN sezione Roma {\it Tor Vergata}, I-00133 Rome, Italy}

\author{Antonino Marcian\`o}
\ead{marciano@fudan.edu.cn}
\address{Department of Physics and Center for Field Theory and Particle Physics, Fudan University, 200433 Shanghai, China}
\address{Laboratori Nazionali di Frascati INFN, Frascati (Rome), Italy, EU}

\date{\today}

\begin{abstract}
We inspect the possibility that neutron star interiors are a mixture of ordinary matter and mirror dark matter.  This is a scenario that can be naturally envisaged according to well studied accretion mechanisms, including the Bondi-Hoyle one. We show that the inclusion of mirror dark matter in neutron star models lowers the maximum  neutron star mass for a given equation of state, and that it decreases the tidal deformability of a given neutron star.  These general features imply that, given an equation of state, one can constrain the maximum viable amount of mirror dark matter in neutron stars in order to consistently fulfill existing maximum mass and tidal deformability constraints.  Conversely, using tidal deformability measurements to rule out equations of state requires making assumptions on the amount of mirror dark matter contained in neutron stars.
Finally, the presence of mirror dark matter also modifies the universal relation that links the tidal deformability of a neutron star to its compactness. Therefore, caution is mandatory when considering exotic models, such as the ones discussed in this paper.
\end{abstract}

\maketitle

\section{\label{Introduction}Introduction\protect\\}
Neutron stars (NSs) are unique natural laboratories to probe the physics of nuclear matter at supranuclear densities \cite{GW1,eos01}.  The behaviour of matter in their interiors is governed by the currently unknown NS equation of state (EoS), which provides a relation between the thermodynamical variables of NS matter --- at the very least between pressure and density.  With terrestrial laboratory experiments, we can test the EoS at a density near the saturation density of nuclei, $2.7\times10^{14} {\rm g}/{\rm cm}^3$, but we cannot reach the density of a NS core, nor deal with the huge number of nucleons that is typical of NSs.  The EoS has an impact on the macroscopic properties of NSs, such as the radius at a given mass, the maximum mass that can be achieved, and the moment of inertia at a given mass \cite{GW3_13}.  Last but not the least, the EoS has a direct impact on how NSs deform under the effect of external tidal fields.  As shown by \citet{FlanaganHinderer}, the NS tidal deformation leaves a clean imprint on the gravitational-wave (GW) signal emitted by inspiralling NS binary systems.  This imprint depends on the \emph{tidal deformability parameter}, which is sensitive to the EoS.  
This implies that measurements of such parameter via GW observations can directly constrain the NS EoS.  Indeed, the LIGO-Virgo \cite{AdvLIGO, AdvVirgo} observation of GW170817 yielded a first direct measurement of the tidal deformability parameter \cite{GW1, GW3}, and sparked several studies aiming at constraining the NS EoS, e.g., \cite{Fattoyev:2017jql, Kumar:2019xgp}.  A second tidal deformability measurement came with the GW190425 event \cite{GW190425}, and constraints on the NS EoS may be produced by combining these two events, e.g., \cite{Vivanco:2020jvj}.

In this work, we study exotic models of NSs containing mirror dark matter (MDM) \cite{Blinnikov:1982eh, Blinnikov:1983gh, Khlopov:1989fj, Berezhiani:2003xm, Berezhiani:2005ek, Berezhiani:2018zvs}. We determine the maximum mass of these models and use the framework provided by Flanagan and Hinderer to calculate their tidal deformabilities.  Our work is motivated by the fact that NS models with ordinary matter alone tend to support high maximum masses and tidal deformabilities or low maximum masses and tidal deformabilities, while current observational data indicates that NSs can be very massive, but at the same time not very deformable.  This statement is supported by GW170817 data \cite{GW3, Zhao_2018}, results from the NICER (the Neutron star Interior Composition Explorer) collaboration \cite{NICER_Bogdanov_2019-01, NICER_Bogdanov_2019-02, NICER_Miller_2019} --- which infers masses and radii of NSs from their X-ray emissions --- and radio timing observations of PSR J0348+0432 \cite{fatpulsar} and PSR J0740+6620 \cite{MassiveNS}. Several studies have considered this very tension between nuclear physics and astrophysical observations, e.g., \cite{Biswas:2020puz}, and there have been attempts to alleviate it with the inclusion of dark matter in NSs \cite{FP00fermionicDM, FP01bosonicDM}, but none with MDM so far.  Our two main constraints are, $M_{\rm max}^{\rm obs}=2\,\msun$ as a representative of the highest NS mass ever measured\footnote{The $\sim 68\%$ confidence measurement for PSR J0348+0432 is $(2.01\pm 0.04)\msun$ \cite{fatpulsar}, while the mass of PSR J0740+6620 is $2.14^{+0.10}_{-0.09}\,\msun$ at $\sim 68\%$ credibility \cite{MassiveNS}.} and the dimensionless tidal deformability parameter for a reference $1.4\,\msun$ NS inferred from GW170817 $\Lambda_{1.4}^{\rm obs}=190^{+390}_{-120}$ \cite{GW3}.

This work is organized as follows. In \textit{Sec.}\,\ref{sec:NSs} we provide some basic notions about NSs, including the approach to calculate the NS tidal deformability. In \textit{Sec.}\,\ref{eoschoice} we explore the EoS models used in this paper, and clarify the reason for our choices.  In \textit{Sec.}\,\ref{darkmatter} we introduce the basic features of the dark matter model under scrutiny in this work, and specify the inclusion of this kind of matter in NSs. Finally, in \textit{Sec.}\,\ref{results} we compare our results with the publicly available GW data for GW170817 \cite{Abbott:2019ebz, Trovato:2019liz}.

We work in $G=c=1$ units, unless otherwise noted.

\section{Neutron Stars} \label{sec:NSs}
NSs are the leftovers of the evolution of stars with masses of at least $\sim 8\,M_\odot$, and core masses of at least $\sim 1.4\,M_\odot$ \cite{Rosswog:2007zz}. They are the second most compact objects known in our Universe, after black holes, with central density of order $\mathcal{O}(10^{15})\,{\rm g}/{\rm cm}^3$.  As a consequence, the EoS that governs the microphysics of matter inside NSs is still unknown, and various candidates are present in the literature \cite{loan, eosrev}.  A non-rotating and isotropic NS in hydrostatic and thermodynamic equilibrium can be modelled as a self-gravitating perfect fluid at zero temperature, $T = 0\,$K, in general relativity.  This is achieved by solving the Tolman-Oppenheimer-Volkoff (TOV) equations \cite{Tolman,Opp},
\begin{eqnarray}
\label{mprime}
\frac{dm(r)}{dr} &=& 4\pi \epsilon(r) r^2\,,\\
\label{pder}
\frac{dp(r)}{dr} &=& -\left[p(r)+\epsilon(r)\right]\frac{d\Phi}{dr}\,,\\
\label{phiprime}
\frac{d\Phi(r)}{dr} &=&\frac{1}{2r}\left[1-\frac{2m(r)}{r}\right]^{-1}\left[8\pi r^2 p(r)+\frac{2m(r)}{r}\right]\,,
\end{eqnarray}
which are derived from the Einstein Equations for the most general spherically symmetric line element
\begin{equation}
ds^2=-e^{2\Phi(r)}dt^2+e^{2\Gamma(r)}dr^2+r^2 d\theta^2 +r^2\sin^2(\theta)d\phi^2\,.
\label{bakline}
\end{equation} 
In these equations, $p(r)$ and $\epsilon(r)$ denote the fluid pressure and energy density, respectively, and $\Gamma(r)$ is governed by the relation
\begin{equation}
\label{lambtov}
e^{-2\Gamma(r)}=1-\frac{2m(r)}{r}\,.
\end{equation}
Closing this system of first order differential equations for the four unknowns $m$, $\epsilon$, $p$, and $\Phi$ requires prescribing an EoS.  Obtaining a specific model requires picking a value of the central energy density and integrating the TOV equations up to the surface of the star, where the condition $p(r=R)=0$ is met.  This allows one to determine two macroscopic properties of the NS model: its radius $R$ and its gravitational mass $M=m(r=R)$.  A family of models, may be obtained by repeatedly solving the TOV equations for a given EoS, while varying the central energy density [see \textit{Fig}.\,\ref{fig:allpure} in \textit{Sec}.\,\ref{eoschoice} for two examples of this].

\subsection{Tidal deformability\protect\\  }
Other than the NS mass and radius, a third macroscopic NS property of interest within this paper is the tidal deformability.  Physically, this quantity is introduced by considering a static and spherically symmetric star, plugged into a static quadrupolar tidal field $\varepsilon_{ij}$ and linearising the response of the star to the deforming field by writing
\begin{equation}
Q_{ij}=-\lambda\varepsilon_{ij}\,,
\label{lambdadef}
\end{equation}
where $Q_{ij}$ is the induced quadrupole momentum of the star, and $\lambda$ is the tidal deformability.  This is related to the $\ell=2$ dimensionless tidal Love number $k_2$ by
\begin{equation}
k_{2}=\frac{3}{2}\lambda R^{-5}\,.
\label{kvslamba}
\end{equation}
In this section we summarise the steps that lead to the equations that are necessary to calculate $\lambda$ for a specific NS model.  We refer the reader to, e.g., Refs.~\cite{Hinderer, Hinderer2} for details.

The starting point is the following first order perturbation of the line element in Eq.\,(\ref{bakline}):
\begin{eqnarray}
ds^{2} &=& -e^{2\Phi(r)}[1+H(r)Y_{20}(\theta ,\phi)]dt^2\nonumber\\
\label{perturbedline}
&& +e^{2\Gamma(r)}[1-H(r)Y_{20}(\theta ,\phi)]dr^2\\
&& +r^{2}[1-K(r)Y_{20}(\theta ,\phi)](d\theta^{2}+\sin^{2}\theta d\phi^{2})\,.\nonumber
\end{eqnarray}
The decomposition into spherical harmonics ($Y_{\ell m}$) is truncated at leading order ($\ell=2$) and, without loss of generality, the azimuthal number $m$ is set to zero.  This is suitable to describe the scenario in which a non-rotating NS is subject to the external tidal field generated by a companion in a compact binary, during the early stages of inspiral.  The deformation will be static and axisymmetric around the line connecting the NS to its companion, which constitutes the axis chosen for the spherical harmonic decomposition.  The line element in Eq.\,(\ref{perturbedline}) allows one to perturb, at first order, the left hand side of the Einstein Equations, i.e., the Einstein tensor $G_{\mu\nu}$.  To complete the picture, the following perturbation of the stress-energy tensor, i.e., the right hand side of the Einstein Equations, is prescribed for the perfect fluid of the NS:
 \begin{equation}
\delta T^{\mu}_{\;\;\nu}={\rm diag}\left( \delta \epsilon(r),\delta p(r),\delta p(r),\delta p(r) \right)Y_{20}(\theta ,\phi).
\label{pertT}
\end{equation}

The $\mu \neq \nu$ perturbed Einstein Equations $\delta G^{\mu}_{\;\;\nu}=0$ lead to the expression
\begin{equation}
K'=H'+2 H \Phi '\,,
\label{kprimo}
\end{equation}
where we use a prime to denote a derivative with respect to the radial coordinate $r$, and where for simplicity we dropped the explicit notation for the dependency on $r$. With the two angular parts of the $\mu = \nu$ perturbed Einstein Equations one may instead write
\begin{equation}
\label{deltap}
\delta p = \frac{\delta G^{2}_{\;\;2}+\delta G^{3}_{\;\;3}}{16\pi Y_{20}(\theta ,\phi )}\,,
\end{equation}
which links $\delta p$ to the metric perturbation, via $\delta G^{2}_{\;\;2}$ and $\delta G^{3}_{\;\;3}$.  Finally, we can combine the remaining components of the perturbed Einstein Equations in the following way:
\begin{equation}
\label{eq:dG00_minus_dG11}
\delta G^{0}_{\;\;0}-\delta G^{1}_{\;\;1}=8\pi\, Y_{20}(\theta ,\phi )[\delta\epsilon -\delta p]\,.
\end{equation}
Because of isotropy, we may write $\delta \epsilon = f(p)\delta p$.  Thus, for small variations of $p$, we may further set $f(p)=\frac{d\epsilon}{dp}$, since the relation $ \delta \epsilon =\frac{d\epsilon}{dp}\delta p$ holds generally, including for small changes in the pressure.  The function $f(p)$ is the inverse of the speed of sound squared, which for a fluid is defined as
\begin{equation}
c_{s}^2(p)=\frac{dp}{d\epsilon}\,.
\end{equation}
This provides a link between the internal structure and the EoS. This allows us to rewrite Eq.\,(\ref{eq:dG00_minus_dG11}) as
\begin{equation}
\label{forh}
\delta G^{0}_{\;\;0}-\delta G^{1}_{\;\;1}=\frac{f(p)-1}{2}\left(\delta G^{2}_{\;\;2}+\delta G^{3}_{\;\;3}\right)\,.
\end{equation}

Substituting the explicit expressions for the perturbed Einstein tensor in Eq.\,(\ref{forh}) and using Eq.\,(\ref{kprimo}) leads to a second order ordinary differential equation for the metric perturbation $H(r)$, which reads
\begin{eqnarray}
&&H(r)\biggl(-\frac{f(p) \Gamma '(r)}{r}-\frac{f(p) \Phi '(r)}{r}-2 \Gamma '(r) \Phi '(r)+\frac{3 \Gamma '(r)}{r}\\
&&-\frac{6 e^{2 \Gamma (r)}}{r^2}+2 \Phi ''(r) -2 \Phi '\,^2(r)+\frac{7 \Phi '(r)}{r}\biggr)\\
&&+H'(r)\biggl(-\Gamma '(r)+\Phi '(r)+\frac{2}{r}\biggr)+ H''(r)=0\,.
\label{odeforH}
\end{eqnarray}
The derivatives of $\Gamma$ and $\Phi$ may be eliminated using the background equations (\ref{mprime})--(\ref{phiprime}).  The differential equation for $H(r)$ needs to be solved in order to determine the tidal deformability.  This is done by integrating outwards, starting from $H(r_0\ll 1) = a_0r_0^2$.  Outside the star, where all fluid variables vanish, the solution to the differential equation may be expressed in terms of the associated Legendre functions $Q^2_{2}(x)$ and $P^2_{2}(x)$ as 
\begin{equation}
H=c_1Q^2_{2}\left(\frac{r}{M}-1\right)+c_2P^2_{2}\left(\frac{r}{M}-1\right)\,.
\label{hqp}
\end{equation}
The coefficients $c_1$ and $c_2$ are determined by matching the interior and exterior solutions at $r=R$.  This procedure leads to the sought expression for the $\ell =2$ tidal Love number:
\begin{eqnarray}
k_2 &=& \frac{8C^5}{5}(1-2C)^2\left[2+2C(y-1)-y\right]\times\nonumber\\
&&\times\biggl\lbrace2C[6-3y+3C(5y-8)]+4C^3[13-11y+\nonumber\\
&&+C(3y-2)+2C^2(1+y)]+\nonumber\\
&&+3(1-2C)^2[(2-y+2C(y-1)]\ln(1-2C)\biggl\rbrace^{-1}\,,
\label{k2}
\end{eqnarray}
where $C=M/R$ denotes the compactness of the star, while $y=RH'(R)/H(R)$. This equation and Eq.\,(\ref{kvslamba}) allow for the calculation of $\lambda$.

As a final comment, we note that it is often common to work with the dimensionless tidal deformability
\begin{equation}
\Lambda=\frac{2}{3}\frac{k_2}{C^5}
\label{eq:Lambda_def}
\end{equation}
and we will do so in the remainder of this paper.

\section{\label{eoschoice}Equation of State Choices\protect\\  }

In this section we motivate our choices for the two EoSs used in this work, namely, SLy~\cite{SLypaper} and MS1~\cite{MS1paper}.

A condition that every viable EoS candidate must satisfy is to be able to sustain the maximum mass constraint [$M_{\rm max}^{\rm obs}=2\,\msun$, see Introduction].  A second condition that must be satisfied by an EoS is that it yields a dimensionless tidal deformability parameter for a $1.4\,\msun$ NS that is compatible with $\Lambda_{1.4}^{\rm obs}=190^{+390}_{-120}$.  \emph{Within an NS modelling approach}, if an EoS cannot produce a stable configuration that supports the mass measured for PSR J0348+0432 and for PSR J0740+6620 or a tidal deformability that agrees with the GW170817 data, that EoS must be ruled out.

\begin{figure}[t]
\centerline{\includegraphics[width=0.85\textwidth]{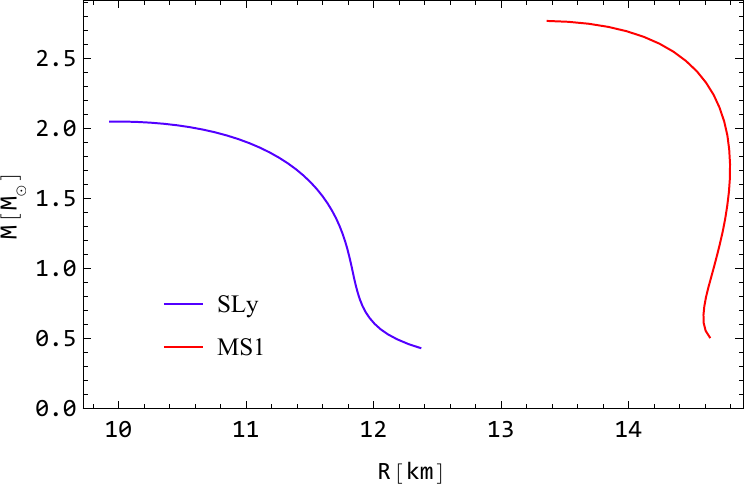}}
\caption{\label{fig:allpure} Neutron star equilibrium sequences obtained with ordinary matter and the SLy (blue) and MS1 (red) equations of state.}
\end{figure}

In \textit{Fig.}\,\ref{fig:allpure}, we show the masses and radii of two NS equilibrium sequences obtained by repeatedly integrating the TOV equations while varying the central density, once assuming the SLy EoS (blue curve) and a second time assuming the MS1 EoS (red curve).  The curves terminate when the compactness, $C=M/R$, reaches its maximum value, beyond which the star becomes unstable.  Both sequences are seen to admit NSs compatible with the maximum mass observational constraint.  MS1 supports NSs with large radii, i.e., low compactness values.  By virtue of Eq.\,(\ref{eq:Lambda_def}), it therefore yields high dimensionless tidal deformabilities and indeed it was found to be incompatible with the GW data of the GW170817 event~\cite{GW1, GW2, GW3}.  Nonetheless, we will show that this incompatibility may be evaded resorting to non-canonical NS models that include MDM, in addition to ordinary matter from the Standard Model.  Essentially, the addition of MDM increases the compactness of a NS with a given gravitational mass. For this very reason, we choose to use MS1 as an EoS in this work.

At the same time, the presence of MDM lowers the maximum NS mass: by inspecting \textit{Fig.}\,\ref{fig:allpure}, one sees that SLy runs the risk of falling short of fulfilling the maximum NS mass constraint when MDM is included in modelling NSs.  The choice of using the SLy EoS therefore enables us to show that there can be a maximum amount of MDM content in NSs.

\section{\label{darkmatter}Mirror Dark Matter\protect\\  } 
Dark matter provides a vast theoretical scenario that encodes the appearance of massive matter fields in the Universe other than the visible ones, in order to fulfill astrophysical constraints. A plausible type of dark matter is MDM \cite{Berezhiani:1995am, Berezhiani:1995yi, MDM00,MDM01}, which arises naturally if one assumes that Nature is parity-symmetric. Asymmetric mirror scenarios have then been deeply analyzed in the literature, with specific focus, for instance, on the possible implications of asymmetric fermionic dark matter for neutron stars \cite{Goldman:2013qla}. The idea behind this model traces back to Lee and Yang's paper on parity violation \cite{LeeYang}. In the same article they proposed a way to restore this symmetry globally, by creating a mirror partner for each particle. This implies the existence of a new sector that is an exact replica of the Standard Model, but with opposite handedness. This symmetry can be interpreted as a spacetime parity that connects each particle $e$, $n$, $p$, $\gamma$ and so on, to the corresponding $e'$, $n'$, $p'$, $\gamma'$, where the prime symbol denotes the mirror sector. Following these assumptions, ordinary and mirror particles will have the same mass and be governed by the same dynamics --- with mirror particles also forming atoms, molecules and astrophysical objects --- but the two sectors will communicate only via gravitational interactions.\footnote{
In a more refined picture, which we do not consider in the present paper, MDM can interact with the Standard Model particles via, for instance, the so called photon--mirror-photon kinetic mixing.  With a mixing of strength of order $\epsilon\sim \mathcal{O}(10^{-9})$, MDM can fulfill all constraints imposed by cosmological observations, including those from the cosmic microwave background and big bang nucleosynthesis~\cite{Foot_2014}.
}
It is also worth mentioning specific predictions for the evolution and structural properties of mirror star massive compact halo objects (MACHOs) \cite{Berezhiani:2005vv}, and the possibility of carrying out tests on mirror neutron oscillations \cite{Berezhiani:2005hv,Berezhiani:2008bc} in forthcoming experiments, including the European Spallation Source (ESS). Bounds on neutron-mirror neutron mixing have been also derived from pulsar timing \cite{Goldman:2019dbq}.

In the scenario of our interest, MDM is tied to galaxy halos, and should mainly appear within the form of cold helium and gases of heavier elements. From the measurements of Bahcall and collaborators~\cite{foot00226}, we can conjecture that dark matter may be found in regions close to those containing stars. Differently, the mass-light ratio would not be constant. These two assumptions point to the perfect environment in which ordinary matter stars can capture dark matter during their lifetime. These are the exact ingredients to apply an accretion criterion such the as the Bondi-Hoyle one~\cite{bondihoyle00,bondihoyle01}.

Another, more sophisticated accretion formula for a body standing in a cloud of gases is derived by X.Y.~Li, F.Y.~Wang and K.S.~Cheng in~\cite{MDMstar01}. Regardless of the specific choice for the accretion model, there is no doubt that the capture of MDM is a phenomenon that must be taken into account in this scenario. The Bondi-Hoyle and the Li-Wang-Cheng accretion models can be seen, respectively, as an upper and a lower limit to the accumulated mass. Assuming that the star is acquiring dark matter, and given a lifetime of one billion years, in both cases the amount of matter acquired exceeds $1\,M_\odot$: this motivates varying the percentage of dark matter mass with respect to the total mass in the range 0\%--50\%.

\subsection{Modelling Neutron Stars Containing Mirror Dark Matter}
\label{sec:2FluidTOV}
The presence of MDM may be incorporated in NS models by treating ordinary baryonic matter ($B$) and MDM as two non-interacting perfect fluids.  This is done by splitting the pressure and energy density that appear in Eqs.\,(\ref{mprime})--(\ref{phiprime}) into two additive contributions:
\begin{eqnarray}
\label{newp}
p(r) &=& p_B(r)+p_{MDM}(r)\,,\\
\label{neweps}
\epsilon(r) &=& \epsilon_B(r)+\epsilon_{MDM}(r)\,.
\end{eqnarray}
In particular, since we are assuming that the two fluids only interact through gravity, Eq.\,(\ref{pder}) separates in two equations:
\begin{eqnarray}
\label{ordpder}
\frac{dp_B(r)}{dr} &=& -\left[p_B(r)+\epsilon_B(r)\right]\frac{d\Phi (r)}{dr},\\
\label{darkpder}
\frac{dp_{MDM}(r)}{dr} &=& -\left[p_{MDM}(r)+\epsilon_{MDM}(r)\right]\frac{d\Phi (r)}{dr}.
\end{eqnarray}
This simplification would not have been possible had we inserted a channel of non-gravitational interaction~\cite{Xiang:2013xwa, MDMstar00, MDMstar01}.

To build NS models that include MDM, one must integrate Eqs.\,(\ref{mprime}), (\ref{phiprime}), (\ref{ordpder}), and (\ref{darkpder}), with the constraints in Eqs.\,(\ref{newp}) and (\ref{neweps}), starting from the center of the star.  To do so, an EoS and a central density need to be specified in both matter sectors.  In principle, we could use two different EoSs, one per matter species \citep[see, e.g.,][]{Leung:2011zz}.  However, it is a good approximation to use the same EoS in both sectors~\cite{MDMstar00, MDMstar04}.  Given the different nucleosyntheses, dark matter nuclei are not distributed like ordinary ones, but it is fair to assume that, during the collapse, the strong gravitational field of the core reduces the nuclei into mirror-protons, mirror-neutrons and small agglomerates of mirror-nucleons that evolve in a way that is similar to that of the ordinary sector, reaching $\beta$-equilibrium.  Therefore, we specify a single EoS that holds in both matter sectors.  As discussed in the previous Section, we adopt the SLy and MS1 EoSs.

The radius $R$ of the NS is determined by integrating the equations of the two-fluid TOV model until the condition that the \emph{total} pressure vanishes is reached, i.e., $p(r=R) = 0$.  The radius of the sphere containing all the baryonic ordinary matter is defined as $R_B\leq R$ such that $p_B(R_B)=0$; similarly, one may define $R_{MDM}\leq R$ such that $p_{MDM}(R_{MDM})=0$ as the radius of the sphere that contains all the MDM present in the NS.

Within the two fluid approach that we outlined so far, it is also possible to define three distinct masses.  The total ordinary matter mass is the integral of $4\pi r^2\epsilon_B(r)$ from $r=0$ to $r=R$, while the total MDM mass is the integral of $4\pi r^2\epsilon_{MDM}$ from the center of the NS up to its surface. Finally, the (total) gravitational mass is instead given by
\begin{equation}
\label{totmass}
M(R)=\int_{0}^{R}4\pi r^2\left[\epsilon_B (r)+\epsilon_{MDM}(r)\right] dr\,.
\end{equation}
This is the quantity we report in our results, unless otherwise stated.

\begin{figure}[!t]
\centerline{\includegraphics[width=0.85\textwidth]{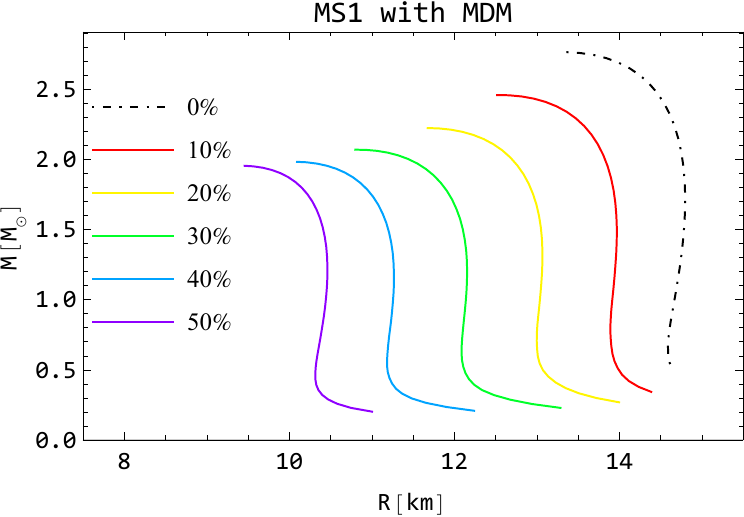}}
\caption{\label{fig:equlibriumseq} NS equilibrium sequences for the MS1 EoS at fix values of the MDM mass to total mass ratio $M_{MDM}/M$.  The legend indicates the value of $M_{MDM}/M$ for a given sequence.}
\end{figure}

\section{\label{results}Results\protect\\}
We now present our results obtained by repeatedly integrating the TOV equations for two fluids (ordinary matter and MDM) described in the previous Section, while varying the total central density and the MDM central density, for a given EoS (which we recall holds for both matter sectors).  This produces a collection of NS models that we interpolate in order to produce NS equilibrium sequences with a fix ratio of MDM mass to total mass, $M_{MDM}/M$.  Because masses are integrated quantities, it is not possible to fix this ratio a priori and an interpolation over a sample of models is necessary. 

In \textit{Fig.}\,\ref{fig:equlibriumseq} we show six NS equilibrium sequences obtained for the MS1 EoS for different values of $M_{MDM}/M$.  The sequence without MDM is shown as a dot-dashed line in the total radius versus total mass plane: in the absence of MDM, the NS radius is maximum at a given NS (total) mass.  As the MDM contribution to the total mass increases, so does the compactness of the NSs, i.e., the radius $R$ decreases for a given $M$.  At the same time, the value of the maximum (total) mass of a stable NS, $M_{\rm max}$, decreases progressively.  Quantitatively, for the MS1 EoS, $R \sim 14.5\,$km and $M_{\rm max}\simeq 2.75\,M_\odot$ in the absence of MDM, while for $M_{MDM}/M = 50$\% $R \sim 10.5\,$km and $M_{\rm max}\simeq 1.95\,M_\odot$.  Since the TOV equations for two fluids that interact only gravitationally are insensitive to swapping the two fluid species, the $M_{MDM}/M = 50$\% sequence collects the most compact configurations that are obtainable for a given EoS.  A sequence with $M_{MDM}/M=90\%$, for example, is identical to a sequence with $M_{MDM}/M = 10\%$.  Indeed, this symmetry was exploited to benchmark the code we wrote to integrate the TOV two-fluid model described in \textit{Sec.}\,\ref{sec:2FluidTOV}.

\begin{table}[!t]
\begin{center}
  \caption{Radius $R$, compactness $C$, Love number $k_2$, and tidal
    deformability $\Lambda$ for $M=1.4M_{\odot}$ NS models built with
    the MS1 EoS (top) and the SLy EoS (bottom), as the percentage of
    MDM mass varies as indicated in the table header.}
  \begin{tabular}{c@{\hspace{0.45cm}}c@{\hspace{0.45cm}}c@{\hspace{0.45cm}}c@{\hspace{0.45cm}}c@{\hspace{0.45cm}}c}
    \addlinespace[0.4em]
    \toprule[1.pt]
    \toprule[1.pt]
    \addlinespace[0.4em]
    $M_{MDM}/M$ & \boldmath$10\%$ & \boldmath$20\%$ & \boldmath$30\%$ & \boldmath$40\%$ & \boldmath$50\%$ \\
    \addlinespace[0.2em]
    \midrule[1.pt]
    \addlinespace[0.2em]
    \multicolumn{1}{l}{R [km]}  & 13.97  & 13.06  & 12.13  & 11.24 & 10.44  \\
    \multicolumn{1}{l}{C}      & 0.148  & 0.158 & 0.170 & 0.184 & 0.198  \\
    \multicolumn{1}{l}{$k_2$ [$10^{-2}$]}  & 8.38 & 7.37 & 6.47 & 6.23 & 6.76 \\
    \multicolumn{1}{l}{$\Lambda$}~ & 786 & 495 & 301 & 197 & 148 \\
    \addlinespace[0.2em]
    \midrule[1.pt]
    \addlinespace[0.2em]
    \multicolumn{1}{l}{R [km]}  & 10.90  & 9.98  & 9.06  & 8.22 & 7.57  \\
    \multicolumn{1}{l}{C}      & 0.190  & 0.207 & 0.228 & 0.251 & 0.273  \\
    \multicolumn{1}{l}{$k_2$ [$10^{-2}$]}  & 6.00 & 4.85 & 4.04 & 3.52 & 3.33 \\
    \multicolumn{1}{l}{$\Lambda$}~ & 163  & 85 & 44 & 23 & 15  \\
    \bottomrule[1.pt]
    \bottomrule[1.pt]
  \end{tabular}
  \label{tab:1.4NSs}
\end{center}
\end{table}

Having seen the general behaviour of a NS equilibrium sequence for a given EoS as we increase the amount of MDM, we report in \textit{Tab.}\,\ref{tab:1.4NSs} the values of the NS radius, compactness, Love number, and tidal deformability ($R$, $C$, $k_2$, and $\lambda$) for a NS with a canonical $M=1.4\,M_\odot$ total mass when varying $M_{MDM}/M$.  These properties are calculated for both the MS1 and the SLy EoSs.

We now turn to a discussion of the general results presented so far, in light of the constraints from GW170817 [$\Lambda_{1.4}^{\rm obs}=190^{+390}_{-120}$], and PSR J0348+0432 and PSR J0740+6620 [$M_{\rm max}^{\rm obs}=2\,\msun$].  The MS1 EoS is not compatible with GW170817 data when considering ordinary matter alone~\cite{GW1}.  However, the inclusion of MDM in an NS model lowers the NS tidal deformability: it is indeed possible to make the MS1 EoS compatible with both the maximum mass constraint \emph{and} the tidal deformability measurement of GW170817, \emph{if} the MDM sector is included when modelling NSs.  In other words, MS1 may be ruled out under the hypothesis that the source of the GW170817 signal contained no MDM, otherwise it cannot.

\begin{figure}[!t]
\centerline{\includegraphics[width=0.85\textwidth]{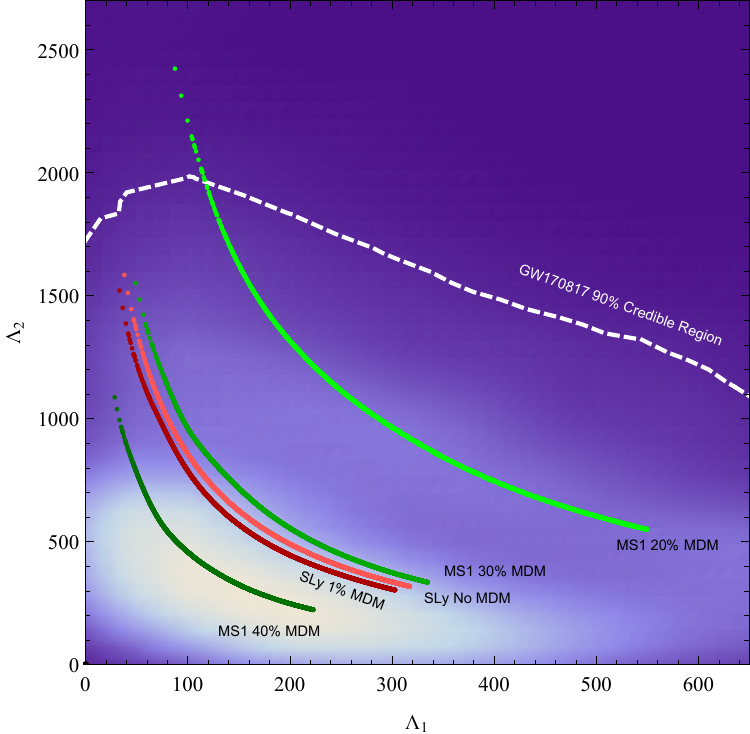}}
\caption{\label{fig:plottone} The background is the two-dimensional posterior distribution of the dimensionless tidal deformabilities of the two NSs in the source of GW170817 \cite{GW1}.  The dashed white line encloses the $90\%$ credible region of such posterior. The colored curves correspond to combinations of EoS and $M_{MDM}/M$ that satisfy the two observational constraints $\Lambda_{1.4}^{\rm obs}=190^{+390}_{-120}$ \emph{and} $M_{\rm max}^{\rm obs}=2\,\msun$: they are obtained by constructing the interpolating function $\Lambda(M)$ and then running it on the component mass posterior of GW170817. The GW170817 data is taken from \cite{datigw}}.
\end{figure}

The case of SLy is different.  Without MDM, this EoS leads to NS models compatible with both the maximum mass constraint and GW170817, but just a small amount of MDM, namely $M_{MDM}/M \simeq 1\%$, can drive the maximum mass below $2\,\msun$.  The presence of MDM may depend on the local environment and, therefore, this statement may be rephrased as follows: under the hypothesis that an SLy-like EoS holds in NS interiors, the amount of MDM in, say, PSR J0348+0432 is bounded by $M_{MDM}/M <1\%$.

All in all, adding MDM as a degree of freedom of NS models alleviates the tension due to $M_{\rm max}^{\rm obs}$ and $\Lambda_{1.4}^{\rm obs}$ discussed in the Introduction.  If one admits the possibility that NSs contain MDM in their interiors, in order for an EoS to meet experimental observations the following criterion --- namely a \emph{necessary} condition --- must be verified: at least one NS equilibrium sequence of the family of sequences yielded by that EoS must satisfy the maximum mass constraint, and at least one sequence must agree with the tidal deformability measurement of GW170817.  Once again, this relies on the fact that the amount of available MDM may depend on the local environment.

\begin{table}[!t]
\centering
\caption{Values of the dimensionless tidal deformability for $1.4\,M_\odot$ NS models built with the the SLy or the MS1 EoS and several values of $M_{MDM}/M$, as denoted in the header of the table. We highlight in bold configurations that are in agreement with both the $\Lambda_{1.4}^{\rm obs}=190^{+390}_{-120}$ \emph{and} the $M_{\rm max}^{\rm obs}=2\,\msun$ observational constraints.}
\begin{tabular}{c@{\hspace{0.45cm}}c@{\hspace{0.45cm}}c@{\hspace{0.45cm}}c@{\hspace{0.45cm}}c@{\hspace{0.45cm}}c@{\hspace{0.45cm}}c}
  \addlinespace[0.4em]
  \toprule[1.pt]
  \toprule[1.pt]
  \addlinespace[0.4em]
  $M_{MDM}/M$ & 0\% & 10\% & 20\% & 30\% & 40\% & 50\% \\
  \addlinespace[0.2em]
  \midrule[1.pt]
  \addlinespace[0.2em]
  \multicolumn{1}{c}{SLy}  & \textbf{282}              & 163           & 85            & 44           & 23     & \multicolumn{1}{c}{15}           \\
  \multicolumn{1}{c}{MS1}    & 1246             & 786           & \textbf{495}            & \textbf{301}           & \textbf{197}            & \multicolumn{1}{c}{148}           \\
  \bottomrule[1.pt]
  \bottomrule[1.pt]
\end{tabular}
\label{tab:adimlam}
\end{table}

Both MS1 and SLy satisfy the necessary condition we just stated. A more interesting criterion is the following \emph{sufficient} condition: for a given EoS to be viable, at least one curve of the family of its NS equilibrium sequences must be in agreement with both the maximum mass constraint and the tidal deformability of GW170817.  A concrete application of this is shown in \textit{Fig.}\,\ref{fig:plottone}.  Here, we plot the two-dimensional posterior distribution for the tidal deformabilities of the two constituents of the GW170817 binary NS (data from \cite{datigw}); the $90$\% credible region for this distribution is delimited by the white dashed contour line.  We also overlay examples of $\Lambda_1$--$\Lambda_2$ curves that satisfy the sufficient criterion just enounced and that were constructed assuming wither the MS1 or the SLy EoS.  Each of these colored curves is obtained by choosing an EoS, fixing the MDM contribution to the total mass $M_{MDM}/M$, constructing the interpolating function $\Lambda(M)$ where $M$ is the NS gravitational mass, and then running this interpolating function on the component mass posterior of GW170817.  This procedure is identical to the one used in \cite{GW1}, but with the additional possibility of including MDM in our NS models.  The EoS and $M_{MDM}/M$ choices label the single curves in the figure.  MS1 satisfies the sufficient condition (i.e., both the tidal deformability and maximum mass constraints) for any value of $M_{MDM}/M$ between $\sim 20\%$ and $\sim 40\%$.  Below this interval, it yields tidal deformability values that are not compatible with GW17087, whereas above this interval it cannot satisfy the maximum mass constraint.  On the other hand, as mentioned previously, SLy is viable on for $M_{MDM}/M \lesssim 1\%$, otherwise it cannot meet the maximum mass constraint given by PSR J0348+0432 and PSR J0740+6620.  These results are summarized in \textit{Tab.}\,\ref{tab:adimlam}, where we report the dimensionless tidal deformability of $1.4\msun$ NSs, $\Lambda_{1.4}$, for both EoSs and various values of $M_{MDM}/M$: configurations that are compatible with $\Lambda_{1.4}^{\rm obs}=190^{+390}_{-120}$ \emph{and} with $M_{\rm max}^{\rm obs}=2\,\msun$ are highlighted in bold.

\begin{figure}[!t]
\centerline{\includegraphics[width=0.85\textwidth]{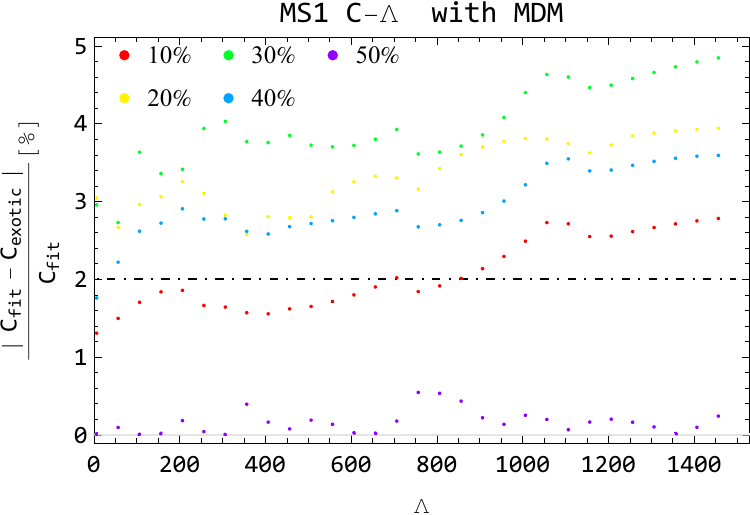}}
\caption{\label{fig:universal} Deviation from the universal behaviour reported in Eq.\,(\ref{clamb}) of the compactness as a function of the dimensionless tidal deformability for NS models that include MDM, for NS masses greater than $1.2\,\msun$.  $C_{\rm fit}$ denotes the compactness yielded by Eq.\,(\ref{clamb}) for a given $\Lambda$, while $C_{\rm exotic}$ is the value obtained in the presence of MDM.  The MS1 EoS is used.  The black dotted-dashed line marks the $2\,\%$ error quoted in \cite{universal} for the universal relation in Eq.\,(\ref{clamb}). The percentages in the legend refer to the ratio $M_{MDM}/M$.}
\end{figure}

\subsection{Universal Relations}
\label{univrel}
Although the EoS that regulates the microphysics of a NS is unknown, some universal, i.e., EoS-independent, relations that connect NS related quantities have be found for canonical matter. For example, \cite{univ01, univ02, Jiang_2020, Gagnon_Bischoff_2018} report an essentially EoS-insensitive relation between the moment of inertia, the tidal Love number, and the spin-induced quadrupole moment of an NS.  Another universal relation expresses, instead, the compactness $C$ of an NS as function of its dimensionless tidal deformability parameter $\Lambda$ \cite{universal}:
\begin{equation}
C=\frac{1}{10}\left[3.71 - 3.91 \cdot 10^{-1}\ln\Lambda + 1.056\cdot (10^{-1}\ln\Lambda)^2\right]\,.
\label{clamb}
\end{equation}
This equation is obtained by computing $\Lambda$ and $C$ for the APR4, MS1, and H4 EoSs, and then fitting the results. The deviations for the fit are quoted to be of at most $\sim 2$\%, and the fits are built for NSs with mass of at least $1.2\,\msun$.

Universal fits like the one above are derived under certain assumptions about NSs. Most commonly, the NSs are taken to be stationary, cold, to have low magnetic fields, etc.  Since this article considers non-canonical NS models that include MDM, we will compare our results for $C$ and $\Lambda$ to the predictions of Eq.\,(\ref{clamb}), in order to assess the deviations from it.  This is interesting as the introduction of MDM breaks one of the assumptions made when deriving this universal relation.  Our results are reported in \textit{Fig.}\,\ref{fig:universal}, where we show the deviation of the compactness of our models from the compactness given by Eq.\,(\ref{clamb}), as a function of $\Lambda$, for different values of $M_{MDM}/M$, and assuming the MS1 EoS.  The $\sim 2$\% deviation quoted for the fit in Eq.\,(\ref{clamb}) is indicated for reference by the dotted-dashed line, and we restrict the NS total mass to be greater than $1.2\,\msun$.  We find that the universal relation in Eq.\,(\ref{clamb}) is indeed followed, with deviations that exceed the $2$\% error only in some extreme cases and by 5\%, at most.  Similar results hold for the SLy EoS.

\begin{figure}[!t]
\centerline{\includegraphics[width=0.85\textwidth]{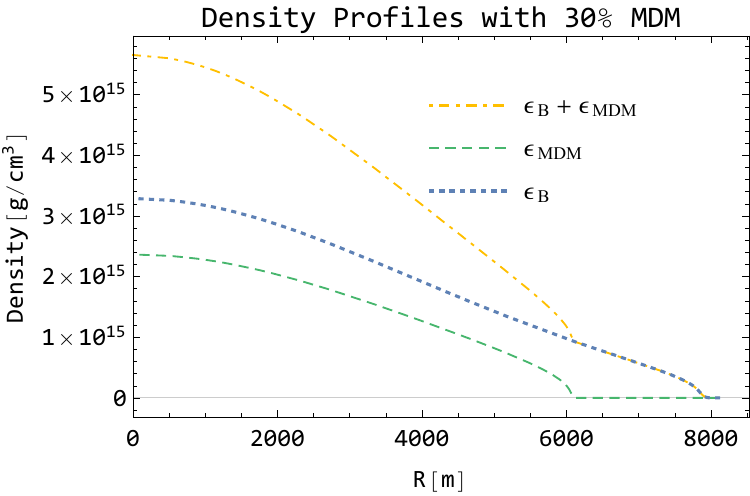}}
\caption{\label{fig:profdens} Energy density profile for a NS model for the SLy EoS and 30\% of MDM.}
\end{figure}

The configurations with the largest deviations from the universal behaviour are those with $M_{MDM}/M = 30\%$.  This is due to how the addition/removal of MDM from the NS models impacts the density profiles of the stars. Since the tidal deformability depends mostly on the proprieties of the outermost layers of the NS \cite{Perot_2020,Piekarewicz_2019}, the deviations from Eq.\,(\ref{clamb}) grow progressively as the distribution of the two fluid species in the outer shell is less and less homogenous.  When MDM is absent or it constitutes all the matter in the NS model, one has a single fluid TOV model, which constitutes the model behind Eq.\,(\ref{clamb}); therefore, deviations from the universal behaviour are negligible.  When $M_{MDM}/M = 50\%$, something similar happens: the two fluids are evenly mixed throughout the star, which effectively behaves as a single fluid TOV model, and again Eq.\,(\ref{clamb}) is verified (purple dots in \textit{Fig.}\,\ref{fig:universal}).  Between $M_{MDM}/M = 0\%$ and $M_{MDM}/M = 50\%$ (and equivalently between $M_{MDM}/M = 50\%$ and $M_{MDM}/M = 100\%$), the two fluids coexist in the inner parts of the NS model but the external shell contains only ordinary matter (equivalently MDM).  A transition layer between the two-fluid and single-fluid region thus exists.  This moves progressively outwards as $M_{MDM}/M$ approaches 50\% (or equivalently inwards as $M_{MDM}/M$ approaches 100\%, starting from 50\%).  An example of this is provided in \textit{Fig}.~\ref{fig:profdens}, where we show the baryonic matter, MDM, and total density profiles of a $1.58\msun$ NS model with $M_{MDM}/M=30\%$, built with the SLy EoS.  The presence of a knee in the total density profile (the two-fluid to single-fluid transition layer) alters the tidal deformability of the star, as one of the two fluids is no longer present. Since the position of this knee depends upon the relative amount of MDM and ordinary matter, the $C$-$\Lambda$ relationship varies with it.  There is therefore an optimal spot between $M_{MDM}/M = 0\%$ and $M_{MDM}/M = 50\%$ (and equivalently between $M_{MDM}/M = 50\%$ and $M_{MDM}/M = 100\%$) that maximises the deviations from Eq.\,(\ref{clamb}): this happens when the ``inohomogeneity'' of the outer layers is greatest.

\section*{\label{conclusions}Conclusions\protect\\  }

Over the last five years, astronomy and astrophysics have been witnessing an epochal revolution: the LIGO-Virgo direct observations of GW signals have opened up the pathway to the exploration of the Universe by means of a new channel of observations.  We have started to probe the Universe according to a new paradigm, going beyond observations performed within traditional astronomy, via electromagnetic radiation.  Additionally, we can combine observational data obtained in multiple observation channels, provided by gravitational and electromagnetic waves, and by neutrinos.

Within this vast panorama of possibilities, our work carried out a systematic study of the equilibrium properties of NSs encompassing MDM. Our research perspective is supported by clues suggesting that dark matter is widespread in galaxies, and that MDM could represent the majority of dark matter in Nature.  Through processes of accretion and capture, this form of dark matter can be assimilated by NSs and modify some of their characteristics, such as the maximum mass, the compactness, and the tidal deformability. To the best of our knowledge, the tidal deformability had never been calculated before for NSs with MDM (but see \cite{FP00fermionicDM} and \cite{FP01bosonicDM} for the cases of fermionic dark matter interacting with ordinary nucleonic matter via the Higgs portal mechanism and of bosonic dark matter with quartic self-coupling, respectively).

The results presented in \textit{Sec.}\,\ref{results} show that NSs with MDM are more compact, less massive and less deformable than canonical NSs, all else being fixed. This circumvents the general tendency of standard NS models to require stiff EoSs in order to support high masses, which comes with the downside of struggling to support low tidal deformabilities, given for example by GW170817 \cite{GW3}.  On the other hand, while soft EoSs can support low tidal deformabilities in canonical NS models, these can struggle to support high mass values.

We carried out a comparison with the dimensionless tidal deformability constraint of GW170817 ($\Lambda_{1.4}^{\rm obs}=190^{+390}_{-120}$ for a $1.4\,\msun$ NS) and the $M^{\rm obs}_{\rm max} = 2\,\msun$ maximum mass constraint of PSR J0348+0432 and PSR J0740+6620. We found that the introduction of MDM in NS models can allow EoSs to satisfy these constraints both separately --- as the local environment of a source and its history affect the amount of MDM a NS contains --- and simultaneously.  Specifically, we showed that the SLy EoS satisfies both constraints if MDM is at most $\sim 1\%$ of the total matter in the star: this implies that if SLy EoS governs NS interiors, then no more than $\sim 1\%$ of the mass of PSR J0348+0432, for example, can be ascribed to MDM. The MS1 EoS, instead, easily satisfies the maximum mass constraint and it can be made compatible with the tidal deformability constraint for percentages of MDM between $20\%$ and $40\%$ of the total NS mass.

Finally, in \textit{Sec}.\,\ref{univrel} we assessed the deviations from the $C$-$\Lambda$ universal behaviour reported in \cite{universal} for standard NSs, when one instead admits the presence of MDM.  We found that in the MDM scenario, for a percentage of MDM $\gtrsim 10\%$ the deviation crosses the $2\%$ tolerance quoted in \cite{universal}.  The difference from the universal trend increases up to about $5$\% until the MDM reaches percentages around $\sim 30\%$ of the total NS mass. At this point, the agreement of the MDM configurations with the universal behaviour improves again. We attribute this behaviour to the distribution of ordinary and MDM near the surface of the NS, as these layers dominate the calculation of $\Lambda$ \cite{Perot_2020,Piekarewicz_2019}.

Future developments of the work carried out in this paper range from testing systematically more EoSs, to exploring other models of dark matter --- for instance, adding a kinetic mix in the case of MDM \cite{Foot_2014}.  With a suite of exotic NS models in hand, comparisons against observational results could then gradually induce restrictions on the parameter space that characterizes viable dark matter candidates. 

\section*{Acknowledgements}
R.C.~and F.P.~thank Andrea Maselli for useful discussions.
F.P.~thanks Bhaskar Biswas for reading the manuscript and providing helpful comments.
The work of A.A. is supported by the Talent Scientific Research Program of College of Physics, Sichuan University, Grant No.1082204112427.
A.M. wishes to acknowledge support by the Shanghai Municipality, through the grant No. KBH1512299, by Fudan University, through the grant No. JJH1512105, and by NSFC, through the grant No. 11875113.
This research has made use of data, software and/or web tools obtained
from the Gravitational Wave Open Science Center
(https://www.gw-openscience.org), a service of LIGO Laboratory, the
LIGO Scientific Collaboration and the Virgo Collaboration. LIGO is
funded by the U.S. National Science Foundation. Virgo is funded by the
French Centre National de Recherche Scientifique (CNRS), the Italian
Istituto Nazionale della Fisica Nucleare (INFN) and the Dutch Nikhef,
with contributions by Polish and Hungarian institutes.

\end{document}